\documentclass{iau}
\usepackage{graphicx,natbib}
 
\title{A 3D view of the Hydra I cluster core\\- II. Stellar populations}

\author[Barbosa et al.]{Carlos Eduardo Barbosa$^{1,2}$\footnote{email: {\tt carlos.barbosa@usp.br}},  
                        Magda Arnaboldi$^1$, Michael Hilker$^1$, 
                        Lodovico Coccato$^1$, Tom Richtler$^3$ \and 
                        Cl\'audia Mendes de Oliveira$^2$}

\affiliation{$^1$European Southern Observatory, Garching, Germany\\
$^2$IAG-USP, S\~ ao Paulo, SP, Brazil\\
$^3$Universidad de Concepci\'on, Concepci\'on, Chile}

\pubyear{2014}
\volume{309}
\jname{Galaxies in 3D across the Universe}
\editors{B. L. Ziegler, F. Combes, H. Dannerbauer, M. Verdugo, eds.}

\begin{document}

\maketitle

\begin{abstract}
Several observations of the central region of the Hydra I galaxy cluster point to a multi-epoch assembly history. Using our novel FORS2/VLT spectroscopic data set, we were able to map the luminosity-weighted age, [Fe/H] and [$\alpha$/Fe] distributions for the stellar populations around the cD galaxy NGC 3311. Our results indicate that the stellar populations follow the trends of the photometric substructures, with distinct properties that may aid to constrain the evolutionary scenarios for the formation of the cluster core.

\end{abstract}

\firstsection
\section{Introduction}

Although previously considered a system in equilibrium, observations of the Hydra I cluster core, dominated by NGC 3311, have shown instead a complex, ongoing process of assembly of matter. Recently, \citet{2012A&A...545A..37A} reported the existence of an faint surface brightness offset envelope. In part I of our work (see Hilker at al. in this proceedings), we have shown that the envelope substructure have line-of-sight velocity distributions distinct from the surrounding regions, which is also related to a group of infalling galaxies. 

\citet{2011A&A...533A.138C} has shown that the stellar content in the region of the offset envelope is different from the other regions, thus attesting that stellar populations can be used to trace and constrain the processes involved in galaxy disruption and in the build up of the massive cD galaxies. However, a spatially resolved study to observe the shape and extent of the stellar population features was still missing. Therefore, here we aim to further demonstrate that the Hydra I core is indeed actively forming using stellar populations as tracers of the past evolution of the system.  

\section{Data and Methods}

We have used our novel FORS2/VLT spectroscopic data set (programme 88.B-448; PI: Richtler), i.e. short slits placed in an onion shell-like pattern onto NGC 3311 to mimic a coarse `IFU'. The data set consists of high resolution optical spectra for 130 positions out to $\sim$30 kpc ($\sim$3 effective radii). We modelled seven absorption
line features in the Lick/IDS system with the alpha-enhanced models of \citet{2011MNRAS.412.2183T} using a Monte Carlo Markov Chains method in order to obtain uncertainties and to access whether we can break the well known metallicity-age degeneracy.

\section{Results and Conclusion}

We present results for the luminosity-weighted ages, metallicities, and alpha element abundances in Fig. 1. The offset envelope exhibits populations which are slightly younger, more metal-rich and have lower alpha abundances. These results are consistent  with the "two stages" formation scenario of central galaxies: the central parts form first "in situ", possibly in a quasi monolithic collapse with violent starbursts. The outer regions then grow by accreting less massive systems with extended periods of star formation.

\begin{figure}
\centering
\includegraphics[width=\linewidth]{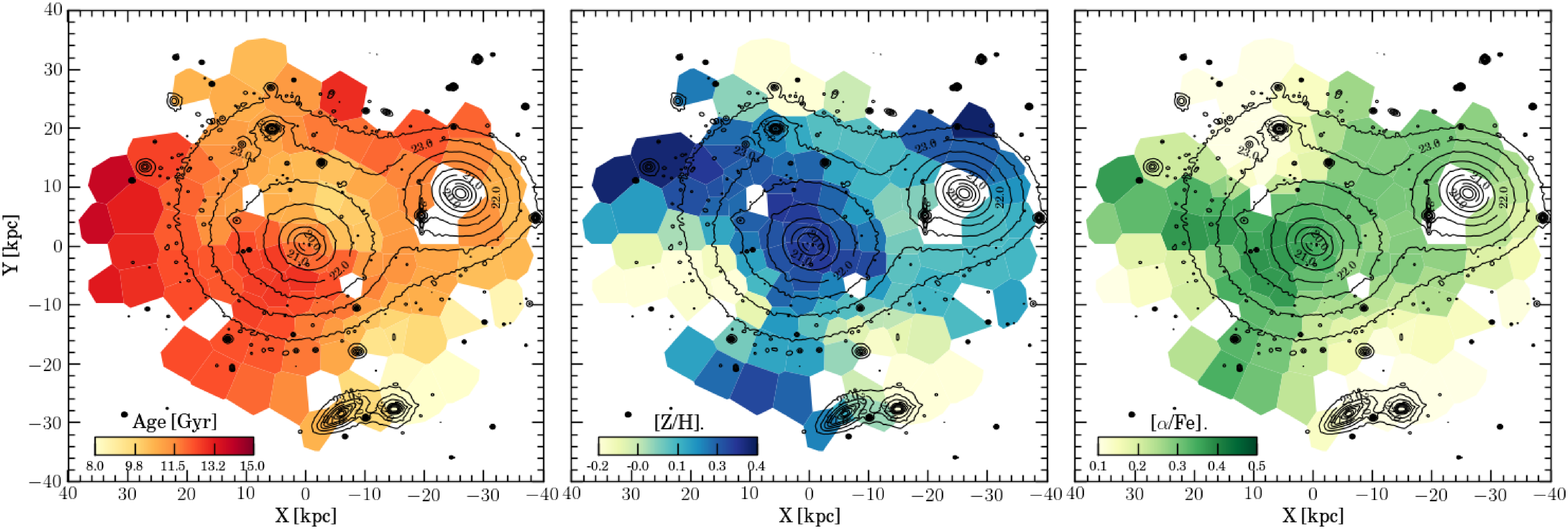} 
\includegraphics[width=\linewidth]{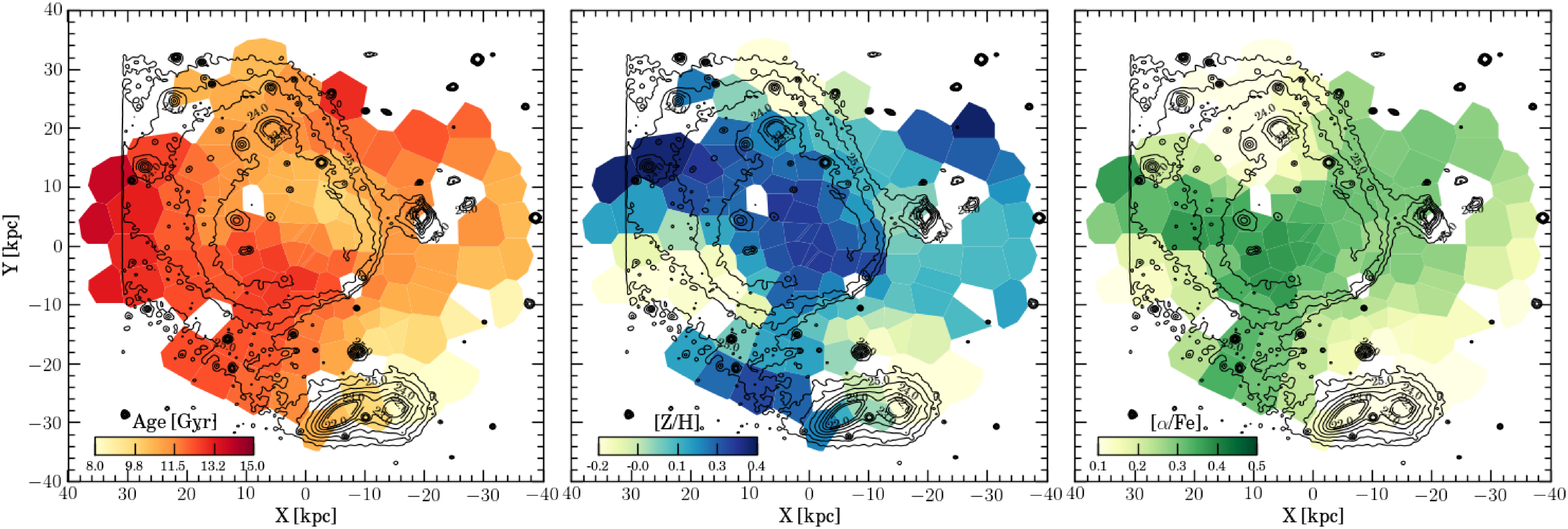} 
\caption{Luminosity-weighted stellar populations of the Hydra I cluster core: ages (left), metallicity (centre) and alpha-elements content (right). Upper and lower panels display contours for the V-band image and the offset envelope respectively, as in \citet{2012A&A...545A..37A}.}\label{fig:fig1}
\end{figure}

\section*{Acknowledgements}

\noindent
Based on observations made with ESO Telescopes at the La Silla Paranal Observatory. CEB and CMdO are grateful  to the S\~{a}o Paulo Research Foundation (FAPESP) 
funding (Procs. 2006/56213-9, 2011/21325-0 and 2012/22676-3). TR acknowledges support from  FONDECYT project Nr.\,1100620, the BASAL Centro de Astrof\'isica y Tecnolog\'ias Afines (CATA) PFB-06/2007, and a  visitorship at ESO/Garching.

\end{document}